\documentstyle[aps,epsfig,12pt]{revtex}

\def\lp {\left( }
\def\rp {\right) }
\def\lb {\left[ }
\def\rb {\right] }
\def\lc {\left\{ }
\def\rc {\right\} }

\def\rar {\rightarrow}
\def\bb {\bibitem}
\def\beq {\begin{equation}}
\def\eeq {\end{equation}}
\def\bea {\begin{eqnarray}}
\def\eea {\end{eqnarray}}
\def\nn {\nonumber}
\def\ni {\noindent}
\def\vs {\vspace}

\def\ao{\~ao }
\def\ii {\'{\i}}
\def\dr {\partial }
\def\di {\partial_\mu }
\def\ds {\partial^\mu }
\def\ub {\bar{u}}

\def\mb {\lambda}
\def\PS {p}
\def\PV {a}
\def\ee {a}
\def\ks {\not\!{k}}
\def\qs {\not\!{q}}
\def\a {\alpha}
\def\b {\beta}
\def\d {\delta}
\def\D {\Delta}
\def\e {\epsilon}
\def\g {\gamma}
\def\m {\mu}
\def\n {\nu}
\def\o {\omega}
\def\p {\pi}
\def\r {\rho}

\def\S {\Sigma }
\def\t {\tau}
\def\th {\theta }
\def\x {\xi }
\def\bD {\mbox{\boldmath $\Delta$}}
\def\f {\mbox{\boldmath $\phi$}}
\def\bk {\mbox{\boldmath $k$}}
\def\br {\mbox{\boldmath $r$}}
\def\bs {\mbox{\boldmath $\sigma$}}
\def\bt {\mbox{\boldmath $\tau$}}

\def\nb {\mbox{\boldmath $\nabla$}}
\def\bx {\mbox{\boldmath $x$}}

\title{THREE-PION EXCHANGE:\\ A GAP IN THE NUCLEON-NUCLEON POTENTIAL}
\author{J.C. Pupin and M.R. Robilotta}
\address{Nuclear Theory and Elementary Particle Phenomenology Group\\ 
Instituto de F\ii sica, Universidade de S\ao Paulo,\\
C.P. 66318, CEP 05315-970, S\ao Paulo, SP, Brazil}

\begin{document}

\maketitle
\vspace{0.5cm}
\begin{abstract}
The leading contribution to the three-pion exchange nucleon-nucleon
potential is calculated in the framework of chiral symmetry. It has
pseudoscalar and axial components and is dominated by the former, which has
a range of about 1.5 fm and tends to enhance the OPEP. The strength of this
force does not depend on the pion mass and hence it survives in the chiral
limit.
\end{abstract}
\vspace{0.5cm}


%111111111111111111111111111111111111111111111111111111111111111111111111111111

\section{Introduction}

The nucleon-nucleon (NN) interaction has been studied for more than 50
years, but it is still not fully understood. The research program proposed
by Japanese physicists around 1950 \cite{J} and based on the idea that the
outer part of the interaction is due to mesonic exchanges proved to be very
successful. Quite generally, the spatial features of a given process are
determined by the mass exchanged in the $t$-channel and lighter systems
correspond to longer interaction ranges. The lightest NN exchange,
associated with the one-pion exchange potential (OPEP) became a consensus in
the sixties, giving rise to a generation of models in which waves with $L>5$
were treated theoretically \cite{OP}.

The second layer of the interaction is much more complex and was studied in
the following two decades, by means of a detailed treatment of the two-pion
exchange potential (TPEP) \cite{P,B}. This process depends on an
intermediate pion-nucleon ($\p$N) scattering amplitude and reflects strongly
its dynamical content.

In the present decade, most of the research work on the NN potential was
aimed at constructing the TPEP in the framework of chiral symmetry. This
symmetry was developed around 1960, for systems in which pions and nucleons
were considered as elementary. Nowadays one believes QCD to be the basic
theory of strong interactions and, accordingly, that pions and nucleons are
made of light quarks, interacting by means of gluon exchanges. As the QCD
Lagrangian predicts that gluons can interact among themselves, calculations
at low and intermediate energies are very difficult. The usual strategy for
overcoming this problem consists in working with effective theories that
treat pions and nucleons as elementary and include, as much as possible, the
main features of the basic theory. The fact that the masses of the u and d
quarks are close to each other and very small in the hadronic scale means
that QCD is approximately invariant under the chiral group SU(2)$\times$%
SU(2). One therefore requires the effective theories to possess approximate
chiral symmetry, besides the usual Poincar\'e invariance.

In low energy processes, chiral symmetry is realized in the Nambu-Goldstone
mode and the vacuum is filled with a condensate, that allows the excitations
of massless collective states, identified with the pions. The breaking of
the symmetry, due to the quark masses at the fundamental level, is
associated with the small pion masses in the effective theories.

Chiral symmetry is very important in the theoretical treatment of two-pion
exchange because it constrains the intermediate $\p$N amplitude. At low and
intermediate energies it is given by a nucleon pole contribution,
superimposed to a smooth background \cite{H83}. The symmetry is responsible
for large cancellations within the nucleon sector that, at once, settle the
scale of the problem and amplify the role of the background. The latter is
very important, since the chiral nucleon sector in isolation does not
suffice for explaining $\pi$N experimental data.

The construction of the TPEP in the framework of chiral dynamics motivated
most of the research of this decade, beginning with the work of Ord\'o\~nez
and van Kolck \cite{OK92}, who considered a system containing just pions and
nucleons. Several works followed, dealing with complementary aspects of the
problem \cite{Ch1,Ch2} and nowadays this part of the NN interaction is well
understood. Predictions for NN observables produced by just the OPEP and the
chiral TPEP, assumed to represent the full interaction for distances larger
than 2 fm, were calculated and shown to agree well with experiment \cite{Ch3}%
. Therefore the effort based on chiral symmetry led to an important
refinement of outer part of this interaction and brought theoretical
constraints to waves with $L\geq 3$.

In the case of the NN interaction, it is important to note that the
importance of chiral symmetry depends strongly on the process one is
considering. In the case of the OPEP, for instance, it is completely
irrelevant, for predictions from chiral $\pi$N Lagrangians coincide with
those arising from interactions without any symmetry. All Lagrangians,
symmetric and non-symmetric, produce exactly the same basic pion-nucleon
vertex, showing that chiral symmetry is compatible with and, at the same
time, irrelevant for the OPEP \cite{R98}.
In the case of the TPEP, on the other hand, the symmetry is crucial. It
produces internal cancellations in the intermediate $\pi$N amplitude which
yield a potential that vanishes in the chiral limit.

To our knowledge, only contributions due to the exchanges of one and two
pions have been so far studied in the framework of chiral symmetry\footnote
{For an early work on problem see, for instance, ref.\cite{HV79}.}.
In order to extend this picture, here we study the component of the NN
potential due to the exchange of three uncorrelated pions. This system has a
mass around 450 MeV and its effects should be longer than those of the vector
mesons usually present in one-boson exchange potentials. The basic interaction
is closely related to the amplitude for the process $\p$N$\rar\p\p$N. Hence,
we review briefly the main features of this reaction in sect.II and
calculate the potential in sect.III.

%222222222222222222222222222222222222222222222222222222222222222222222222222222

\section{Pion Production}

The NN interaction mediated by the exchange of three uncorrelated pions is
derived from $T_{cba}$, the amplitude for the process $\p^a(k) N(p)\rar \p^b
(q)\;\pi ^c (q^{\prime}) N(p^{\prime})$. This amplitude is given by the
diagrams of fig.1 and written as the sum of $T_{cba}^{\p}$, representing the
class of processes with a pion pole in the $t$-channel, and a remainder,
denoted by $\bar{T}_{cba}$, whose dynamical content is indicated in fig.1.
In the framework of chiral symmetry, this last amplitude is given by a basic
family of diagrams, involving only pions and nucleons, supplemented by other
processes, containing deltas, rhos, omega and the $\pi N$ $\sigma $-term.

\begin{figure}[h]
\centerline{\epsfig{figure=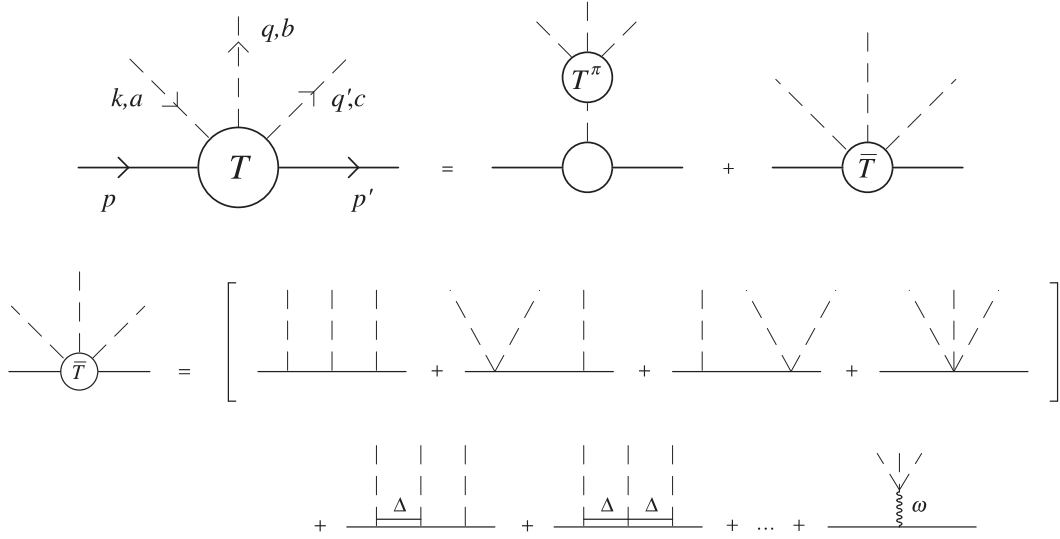,width=14.0cm}}
\caption{Diagram for pion production.}
\end{figure}

There are many alternative ways of implementing chiral symmetry. In
particular, the subset of diagrams in the pure pion-nucleon sector,
corresponding to the minimal realization of chiral symmetry in this problem,
may be evaluated by means of non-linear Lagrangians with $\p$N couplings
that may be either pseudovector (PV) os pseudoscalar (PS). Denoting the pion
field by $\f$, and defining $f= (f_\pi^2-\f^2)^{1/2}$, we have

\bea
{\cal L}_{PV} &=& {\cal L}_\p \!+\! \bar{\psi}\lc i\gamma_\mu \lb \ds%
\! +\! i\frac{\f \times \ds\f}{f_\pi(f\!+\!f_\pi)}\cdot \frac{\bt}{2}\rb%
\!-\!m\rc \psi \!+\! \frac{g_A}{2f_\pi} \bar{\psi}\gamma_\mu \gamma_5 \bt %
\psi\! \cdot\! \lp  \ds\f - \frac{\ds f \f}{f+f_\pi}\rp \;,
\label{2.1}\\[2mm]
{\cal L}_{PS} &=& {\cal L}_\p
+ \bar{N}i\not\!\dr N -g \bar{N}(f+i\bt\cdot\f \gamma_5)N \;,
\label{2.2}
\eea

\ni where

\beq
{\cal L}_\p = \lb \frac{1}{2} (\di f \ds f\!+\! \di\f\!\cdot\!\ds\f)\!+\!
f_\pi\mu^2 f \rb\!\;. \label{2.p} 
\eeq

\ni In these expressions, $\psi$ and $N$ are the nucleon fields with
non-linear and linear transformation properties, $\m$ and $m$ are the pion
and nucleon masses and $f_\p$, $g$ and $g_A$ are respectively the pion
decay, the $\p$N coupling and the axial decay constants. It is important to
stress that these Lagrangians, in spite of their different aspects, have the
same dynamical content and physical results do not depend on the particular
version one adopts, as demonstrated on general grounds \cite{CWZ}.

The pion production amplitude has the general form

\beq
iT_{cba}= \ub \lb \d _{bc}\t_a \;(A^\p+\bar{A}) +\d _{ac}\t _b\;(B^\p+\bar{B}%
) +\d _{ab}\t_c\;(C^\p+\bar{C})+ i\e_{cba}\;\bar{E} \rb\!\g _5\; u\;, \label%
{2.3} 
\eeq

\ni where the tags $\p$ and $bar$ refer to the pion-pole and background
contributions. At threshold, this amplitude is usually written as

\beq
\left. 
iT_{cba}\rb^{th}= 2m\;\bs\!\cdot\!\bk \lb D_1 \lp \d_{ac}\t _b+\d _{ba%
}\t _c\rp +D_2\;\d _{cb}\t_a\rb\;, \label{2.4} 
\eeq

\ni where $D_1$ and $D_2$ are dynamical coefficients. Their empirical values
may be obtained from the following specific processes

\bea
(\p^- p\rar \p^+ \p^- n) \rar \left. iT\rb^{th} &=& 2\sqrt{2}\;m\; \bs%
\!\cdot\!\bk\;D_1 \;, \label{2.5} \\
(\p^+ p\rar \p^+ \p^+ n) \rar \left. iT\rb^{th} &=& \sqrt{2}\;m\; \bs%
\!\cdot\!\bk\;\lp D_1+D_2 \rp \;. \label{2.6} 
\eea

The pion-pole amplitude for on-shell nucleons is

\beq
iT_{cba}^\p = -\;\frac{mg_A}{f_\p}\; \lb \ub\;\t _d\;\g_5\; u \rb
\frac{T_{dcba}^{\p\p}}{(p'\!-\!p)^2-\m^2}\; , \label{2.7} 
\eeq

\ni where $T_{dcba}^{\p\p}$ is the pion scattering amplitude. At tree level,
it is given by

\beq
T_{dcba}^{\p\p} = \frac{1}{f_\p^2}\; \lc \d _{ad}\d _{bc}\lb %
(q\!+\!q')^2\!-\!\m^2\rb
\!+\!\d_{bd}\d _{ac}\lb (k\!-\!q')^2\!-\!\m^2 \rb
\!+\!\d_{cd}\d _{ab}\lb (k\!-\!q) ^2\!-\!\m^2 \rb \rc
\label{2.8} 
\eeq

\ni and yields

\beq
A^\p = -\; \frac{mg_A}{f_\p^3}\;\frac{(p'\!-\!p\!-\!k)^2-\m^2}{(p'\!-\!p)^2-%
\m^2}\;. \label{2.9} 
\eeq

The contributions to $\bar{A}$, calculated with the PV Lagrangian, are given
by

\bea
\bar{A} &=& \lp \frac{mg_A}{f_\p}\rp^3 \lc\frac{\qs' \ks}{(
q'^2\!+\!2p'\!\cdot\!q') ( k^2\!+\!2p\!\cdot\! k) } +\frac{\ks\qs'}{(
k^2\!-\!2p'\!\cdot\! k) (q'^2\!-\!2p\!\cdot\! q')} + \frac{\qs\ks}{(
q^2\!+\!2p'\!\cdot\! q) ( k^2\!+\!2p\!\cdot\! k) } \right. \nn\\[2mm]
&+& \left. \frac{\ks\qs}{(k^2\!-\!2p'\!\cdot\! k) ( q^2\!-\!2p\!\cdot\! q)} +%
\frac{\qs'\qs}{( q'^2\!+\!2p'\!\cdot\! q') (q^2\!-\!2p\!\cdot\! q)} +\frac{%
\qs\qs'}{( q^2\!+\!2p'\!\cdot\! q) (q'^2\!-\!2p\!\cdot\! q')} \right. \nn\\[%
2mm]
&-&\left.\;\frac{1}{m} \lp\frac{\ks}{k^2\!+\!2p\!\cdot\! k}+ \frac{\ks}{%
k^2\!-\!2p'\!\cdot\! k}\rp 
\right. \nn\\[2mm]
&+& \left. \frac{1-1/g_A^2}{4m^2} \lb -2\;\frac{\ks}{m} +\frac{(\ks\!+\!\qs')%
\qs}{q^2\!-\!2p\!\cdot\!q}+\frac{(\ks\!+\!\qs)\qs'}{q'^2\!-\!2p\!\cdot\!q'} +%
\frac{\qs (\ks\!+\!\qs')}{q^2\!+\!2p'\!\cdot\!q}+\frac{\qs'(\ks\!+\!\qs)}{%
q'^2\!+\!2p'\!\cdot\!q'} \rb \rc \;, \label{2.10} 
\eea

\ni the corresponding expressions for $B$ and $C$ are obtained
by making $k\rar-q$ and $k\rar-q^{\prime }$ respectively, and $\bar{E}$ is

\bea
\bar{E} &=& \lp \frac{mg_A}{f_\p}\rp^3 \lc
\frac{\qs'\ks}{( q'^2\!+\!2\,p'\!\cdot\! q') ( k^2\!+\!2p\!\cdot\! k)} -%
\frac{\ks\qs'}{( k^2\!-\!2p'\!\cdot\! k) (q'^2\!-\!2p\!\cdot\! q')} -\frac{%
\qs\ks}{( q^2\!+\!2p'\!\cdot\! q) ( k^2\!+\!2p\!\cdot\! k) } \right. \nn\\[%
2mm]
&+& \left. \frac{\ks \qs}{(k^2\!-\!2p'\!\cdot\! k) ( q^2\!-\!2p\!\cdot\! q) }
+ \frac{\qs'\qs}{( q'^2\!+\!2p'\!\cdot\! q') ( q^2\!-\!2p\!\cdot\! q) } -%
\frac{\qs \qs'}{( q^2\!+\!2p'\!\cdot\! q) ( q'^2\!-\!2p\!\cdot\! q') }\rc %
\;. \label{2.11} 
\eea

For future purposes, one notes that if the PS Lagrangian were used, one
would obtain the same structure with $g_A=1$ and the last line of the eq.
for $\bar{A}$ would vanish. In the PS case, the signature of chiral symmetry
are the contact interactions due to the function $f$ in eq.(\ref{2.2}),
which give rise to the terms proportional to $1/m$ in eq.(\ref{2.10}).

In order to estimate the accuracy of the pion-production amplitude derived
from eq.(\ref{2.1}), we consider the contributions to the amplitudes $D_1$
and $D_2$, which are written in terms of the variables

\bea
D_1^\p &=& \frac{g_A}{2f_\p^3}\;\sqrt{\frac{2m}{E\!+\!m}} \;\frac{\m\; (
2\o\!-\!\m) }{2m( E\!-\!m)\!+\!\m^2}\;, \label{2.12}\\[2mm]
\bar{D}_1 &=& -\frac{g_A^3}{4 f_\p^3}\sqrt{\frac{2m}{E\!+\!m}} \lb 
\frac{2m}{2m\!+\!\m} \! \lp 1\!- \frac{\m}{2E\!-\!\m}\rp
\! - \! \lp 1\!- \frac{1}{g_A^2}\rp \! \lp 1\!- \frac{\m}{m}
-\frac{\m}{2E\!-\!\m}+\frac{\m}{2m\!+\!\m}\rp\rb , \label{2.13} 
\eea

\ni and

\bea
D_2^\p &=& -\; \frac{g_A}{2f_\p^3}\;\sqrt{\frac{2m}{E\!+\!m}}\; \frac{3\m ^2%
}{2m (E\!-\!m)\!+\!\m^2}\;, \label{2.14}\\[2mm]
\bar{D}_{2} &=&-\;\frac{g_A^3}{4 f_\p^3}\sqrt{\frac{2m}{E+m}}\, \lb \frac{%
4m^2}{2m\o\!-\!\m^2}\lp \frac{\m}{m}+\frac{\m}{2E\!-\!\m}\rp
-\frac{m}{2m\!+\!\m}\lp 1+\frac{4m}{2E\!-\!\m}\rp
\right. \nn\\[2mm]
&-& \left. 2 \lp 1-\frac{1}{g_A^2}\rp \lp \frac{\m}{m}+\frac{\m}{2E\!-\!\m}-%
\frac{\m}{2m\!+\!\m}\rp \rb \;, \label{2.15} 
\eea

\ni where $\o=[\m ( 4m\!+\!5\m)]/[2( m\!+\!2\m)]$ and $E = m\!+\!2\m\!-\!\o$%
. Expanding these amplitudes in powers of $\mu /m$, one has

\bea
D_1 &=& [D_1^\p]+[\bar{D}_1] \simeq \lb \frac{g_A}{8f_\p^3} \;\lp 3+ \frac{3%
}{2}\; \frac{\m}{m} + \cdots \rp\rb
-\lb \frac{g_A}{8 f_\p^3}\lp 2-2 \;\frac{\m}{m} +\cdots \rp \rb \;, \label%
{2.16} \\
\nn\\[2mm]
D_2 &=& [D_2^\p]+[\bar{D}_2] \simeq -\lb \frac{g_A}{8 f_\p^3} \lp 3+ \frac{9%
}{2}\; \frac{\m}{m} +\cdots \rp \rb
-\lb \frac{g_A}{8 f_\p^3} \lp 4\; \frac{\m}{m} +\cdots \rp\rb \;. \label%
{2.17} 
\eea

\bigskip

The results for the full amplitudes at threshold, namely $D_1= g_A (1+7\m%
/2m)/8f_\p^3$ and $D_2= -g_A (3+17\m/2m)/8f_\p^3$, coincide with those
obtained in the framework of chiral perturbation theory \cite{BKM}. In order
to asses the role of chiral symmetry in this problem we note that a
Lagrangian without any symmetry, containing just a PS $\p$N interaction,
would give rise to the same pion-pole contribution and an amplitude $\bar{A}$
corresponding to just the six first terms in eq.(\ref{2.10}), which involve
two nucleon propagators. Therefore the full elimination of chiral symmetry
would yield $\bar{D}_1\rar g_A(2)/8f_\p^3$ and $\bar{D}_2\rar - g_A(-4+3\m%
/m)/8f_\p^3$, indicating that chiral symmetry does play a role in this
problem. On the other hand, that this role is not as large as in the case of 
$\p$N$\rar\p$N, where the same procedure would change one of the scattering
lenghts by a factor of 200. Numerical results for the amplitudes are given
in table~\ref{Tab.1} and show that predictions from the minimal chiral model
are close to empirical values, although there is some room for improvement in
$D_2$.

\bigskip \bigskip \bigskip

\begin{table}
\caption{Subamplitudes $D_{1}$ and $D_{2}$ in $\m^{-3}$ units. Experimental
results correspond to a best fit quoted in ref.[12].}
\bigskip
\begin{tabular}{ccccccc}
& $D_1^\p$ & $\bar{D}_1$ & $D_1$ & $D_2^\p$ & $\bar{D}_2$ & $D_2$ \\ \hline
eqs.($\ref{2.12}-\ref{2.15}$) & 1.78 & -0.91 & 0.87 & -2.02 & -0.30 & -2.32
\\
CHPT, eqs.(\ref{2.16},\ref{2.17}) & 1.81 & -0.96 & 0.85 & -2.06 & -0.33 & 
-2.39 \\
experiment & - & - & 0.80 & - & - & -3.20 
\end{tabular}
\label{Tab.1}
\end{table}

In this work we are interested in the construction of the NN interaction due
to the exchange of three pions, which is based on the amplitude $\bar{T}
_{cba}$. As indicated in fig.3, the complete evaluation of this amplitude
would require the calculation of a large number of diagrams. However, long
ago Olsson and Turner \cite{OT} have shown that the leading contribution to
this amplitude comes from the effective Lagrangian

\beq
\bar{\cal{L}} = \frac{g_A}{8f_\pi^3} \; \bar{\psi} \g _\m\;\g _5 \bt\;
\psi\!\cdot\!\f\; \ds\f^2 \;. \label{2.19} 
\eeq

\ni It gives rise to the following contribution to $\bar{A}$

\beq
\bar{A}= \frac{2g_A}{8f_\p^3} \lp 2m-\ks \rp\; \label{2.20} 
\eeq

\ni and, as before, $\bar{B}$ and $\bar{C}$ are obtained by making $k\rar -q$
and $k\rar -q^{\prime}$ respectively. This corresponds to the threshold
amplitudes

\bea
\bar{D}_1&=& -\; \frac{g_A}{8f_\p^3}\; \sqrt{\frac{2m}{E\!+\!m}}\; \lp 2+%
\frac{\mu }{m}\rp
\simeq -\; \frac{g_A}{8f_\p^3} \lp 2+\frac{\mu }{m}\rp \;, \label{20.21}\\[%
2mm]
\bar{D}_2 &=& \; \frac{g_A}{8f_\p^3}\;\sqrt{\frac{2m}{E\!+\!m}}\; \lp \frac{2%
\m }{m}\rp
\simeq \frac{g_A}{8f_\p^3}\lp\frac{2\m }{m}\rp \;, \label{20.22} 
\eea

\ni showing that the effective Lagrangian given by eq.(\ref{2.19})
reproduces correctly the leading contribution at threshold.

%333333333333333333333333333333333333333333333333333333333333333333333333333333

\section{Nucleon-Nucleon Interaction}

The basic element in the construction of the NN potential due to the
exchange of three uncorrelated pions (3PEP) is the corresponding Born
amplitude for the process
N$(p_{1})$N$(p_{2})\rar$N$(p_{1}^{\prime })$N$(p_{2}^{\prime })$,
associated with the diagrams of figure 2. Denoting this
amplitude by $F$, one has

\beq
F= \frac{1}{3!}\int \frac{d^4 Q}{(2\p )^4}\int \frac{d^4 Q'}{(2\p)^4}\;\D %
(k)\;\D (q)\;\D (q')\;\;\bar{T}_{cba}^{(1)}\;\;\bar{T}_{cba}^{(2)}\;, \label%
{3.1} 
\eeq

\ni where the factor $1/3!$ is due to the symmetry of the intermediate
three-pion state, $\D $ is a pion propagator and $\bar{T}_{cba}^{(i)}$ is
the pion production amplitude for nucleon $i$.

\begin{figure}[h]
\centerline{\epsfig{figure=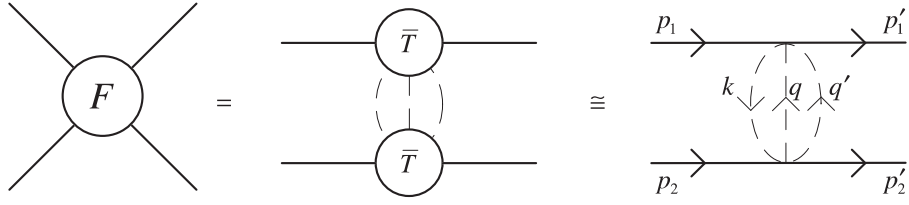,width=12.0cm}}
\caption{Leading contribution to the three-pion exchange potential.}
\end{figure}

We adopt the following external kinematic variables

\bea
W &=& p_1+p_2 = p'_1 + p'_2 \;, \label{3.2} \\
\D &=& p'_1-p_1 = p_2-p'_2 \;, \label{3.3} \\
z &=& [ ( p_1 + p'_1 ) - ( p_2 + p'_2 ) ]/2\;. \label{3.4} 
\eea

\ni As the nucleons are assumed to be on-shell, they are constrained by

\beq
W \!\cdot\! z=W\!\cdot\! \D = z\!\cdot\!\D = 0 \;. \label{3.5} 
\eeq

For the internal variables we define

\bea
Q &=& (q+q'+k) /2 \;, \label{3.6} \\
Q' &=& (q'-q) /2\;, \label{3.7} 
\eea

\ni so that

\bea
k&=& Q-\D/2\;, \label{3.8} \\
q&=& Q/2+\D/4-Q'\;, \label{3.9}\\
q'&=& Q/2+\D/4+Q' \label{3.10} 
\eea

\ni and the condition of momentum conservation reads $q+q^{\prime}-k=\D$.

As discussed in the previous section, the leading contribution to the
amplitude $\bar{T}$ comes from the effective Lagrangian given by eq.(\ref
{2.19}), which yields the following intermediate effective vertex for
nucleon (2)

\beq
\bar{T}_{cba}^{(2)} = i \lp\frac{g_A}{4f_\p^3}\rp \lp \ub \g^\m\g_5 u \rp%
^{(2)} \lb \d_{bc}\t_a^{(2)}\; (q+q')_\m + \d_{ac}\t_b^{(2)}\; (q'-k)_\m +
\d_{ab}\t_c^{(2)}\; (q-k)_\m \rb \;. \label{3.11} 
\eeq

\ni The corresponding expression for nucleon (1) has the same form, but is
globally multiplied by (-1), due to the senses of flow of internal momenta.
Using these results into eq.(\ref{3.1}), we have

\beq
F(\D) = \lp\frac{ g_A}{4f_\p^3}\rp^2 \; \bt^{(1)}\!\cdot\!\bt^{(2)}\; \lp %
\ub \g^\m\g_5 u \rp^{(1)}\lp \ub \g^\n\g_5 u \rp^{(2)}\;I_{\m\n}(\D) \;, 
\label{3.12} 
\eeq

\ni where $I_{\m\n}$ is given by

\bea
I_{\m\n}(\D) &=& \frac{1}{3!}\; \int \frac{d^4 Q}{(2\p )^4}\; \frac{1}{%
[(Q\!-\!\D/2)^2\!-\!\m^2]} \nn\\[2mm]
&\times& \int \frac{d^4 Q'}{(2\p)^4}\; \frac{[3Q_\m Q_\n\!-\!Q_\m\D_\n/2\!-\!%
\D_\m Q_\n/2\!+\!27\D_\m\D_\n /4+4Q'_\m Q'_\n]} {[(Q'\!-\!Q/2\!-\!\D%
/4)^2\!-\!\m^2][(Q'\!+\!Q/2\!+\!\D/4)^2\!-\!\m^2]}\;. \label{3.13} 
\eea

This function is evaluated in the appendix and reads

\beq
I_{\m\n}= \D_\m\D_\n \; \m^4\;I^\PS(\D) + g_{\m\n}\; \m^6 \;I^\PV(\D) \;, 
\label{3.14} 
\eeq

\ni with $I^\PS$ and $I^\PV$ given by eqs.(\ref{a.12}) and (\ref{a.13})
respectively. Using the nucleon equation of motion, one has

\bea
F(\D) = - \lp \frac{ \;g_A}{4 f_\p^3}\rp^2 \bt^{(1)}\!\cdot\bt^{(2)} &&\lb\;
4m^2 \m^4\; \lp \ub \g_5 u \rp^{(1)}\lp \ub \g_5 u \rp^{(2)}\;I^\PS(\D)
\right. \nn\\
&&- \left. \m^6 \; \lp \ub \g^\m\g_5 u \rp^{(1)}\lp \ub \g_\m\g_5 u \rp%
^{(2)}\;I^\PV(\D) \rb \;. \label{3.15} 
\eea

This expression corresponds to the exchanges of pseudoscalar and axial
systems. In order to make the strength of the interaction more transparent,
we eliminate $g_A$ in favour of $g$, by means of the G-T relation, $g_A = g
f_\p /m$, and write

\bea
F(\D) = -\; g^2\; \bt^{(1)}\!\cdot\bt^{(2)} \lp\frac{\m^2}{2 f_\p^2}\rp^2 && %
\lb \lp \ub \g_5 u \rp^{(1)}\lp \ub \g_5 u \rp^{(2)}\;I^\PS(\D) \right. \nn%
\\
&&-\left. \frac{\m^2}{4m^2} \lp \ub \g^\m\g_5 u \rp^{(1)}\lp \ub \g_\m\g_5 u %
\rp^{(2)}\;I^\PV(\D) \rb  \;. \label{3.16} 
\eea

For future purposes, it is worth noting that the corresponding amplitude for
one-pion exchange is

\beq
F^\p(\D) = -\; g^2 \; \bt^{(1)}\!\cdot\bt^{(2)} \lb \lp \ub \g_5 u \rp^{(1)}%
\lp \ub \g_5 u \rp^{(2)}\;\frac{1}{\D^2-\m^2}\rb \;. \label{3.15b} 
\eeq

Going to the non-relativistic limit in the center of mass frame, one has

\bea
\frac{F(\bD)}{4m^2}\rar f(\bD) = - \lp\frac{g}{2m}\rp^2 \bt^{(1)}\!\cdot\bt%
^{(2)} \lp \frac{\m^2 }{2 f_\p^2}\rp^2 && \lb - \bs^{(1)}\!\cdot\!\bD\;\bs%
^{(2)}\!\cdot\!\bD \;I^\PS(\D) \right. \nn\\
&&+ \left. \m^2 \; \bs^{(1)}\!\cdot \!\bs^{(2)}\; I^\PV(\D) \rb \;. \label%
{3.17} 
\eea

This result allows the configuration space potential to be written as

\bea
V^{3\p}(x) &=& -\;\frac{\m}{4\p}\; \int \frac{d^3\bD}{(2\p)^3} \;e^{-i \bD%
\cdot\br}\lb \frac{4\p}{\m}\; f(\bD) \rb \nn \\[2mm]
&=& \lp\frac{g\m}{2m}\;\frac{\m^2 }{2 f_\p^2} \rp^2 \frac{\m}{4\p}\bt%
^{(1)}\!\cdot\bt^{(2)} \lb  \bs^{(1)}\!\cdot\!\nb_x \;\bs^{(2)}\!\cdot\!\nb%
_x \;U^\PS (x) + \bs^{(1)}\!\cdot \!\bs^{(2)}\;U^\PV (x) \rb\;, \label{3.18}
\eea

\ni where the $U(x)$ are integrals of Yukawa functions, written in terms of
the variable $x\equiv \m r$ and given by eqs.(\ref{a.19}) and (\ref{a.20}).

Using the result

\bea
\bs^{(1)}\!\cdot\!\nb_x \bs^{(2)}\!\cdot\!\nb_x &&\lb \lp1+\frac{3}{\ee x}+%
\frac{3}{\ee^2x^2}\rp \frac{e^{-\ee x}}{x^3}\rb 
= \frac{\ee^2}{3} \lb \bs^{(1)}\!\cdot\bs^{(2)} \lp 1+\frac{7}{\ee x}+\frac{%
27}{\ee^2x^2}+\frac{60}{\ee^3x^3}+\frac{60}{\ee^4x^4}\rp  
\right. \nn\\[2mm]
&&+ \left. S_{12} \lp 1+ \frac{10}{\ee x}+ \frac{45 }{\ee^2 x^2}+\frac{105}{%
\ee^3x^3}+\frac{105}{\ee^4 x^4} \rp 
\rb \frac{e^{-\ee x}}{x^3} \;, \label{3.21} 
\eea

\ni with $S_{12}= 3 \bs^{(1)}\!\cdot\! \hat{\bx}\; \bs^{(2)}\!\cdot\! \hat{%
\bx} - \bs^{(1)}\!\cdot\! \bs^{(2)}$, into eqs.(\ref{a.19}) and (\ref{a.20}%
), we obtain

\bea
V^{3\p}(r) &=& \frac{1}{3} \lp\frac{g\m^3}{4m f_\p^2}\rp^2 \frac{\m}{4\p}\; %
\bt^{(1)}\!\cdot\bt^{(2)} \lc \bs^{(1)}\!\cdot\!\bs^{(2)} \lb  U_0^\PS (x) +
3\; U^\PV(x) \rb
+S_{12}\; U_2^\PS (x) \rc \;, \label{3.22} 
\eea

\ni where

\bea
U_0^\PS(x) &=& \frac{1}{(4\p)^4}\; \frac{1}{6}\int_0^1\!d\a \int_0^1\!d\g \; %
\g\;\ee^4\; \lc  [3\!+\!(1\!-\!2\a)^2](1\!-\!2\g)^2 + 2 [1\!-\!(1\!-\!2\a%
)^2] (1\!-\!2\g) \right. \nn\\[2mm]
&+&\left.  [27\!+\!(1\!-\!2\a)^2]\rc
\lp 1+\frac{7}{\ee x}+\frac{27}{\ee^2x^2}+\frac{60}{\ee^3x^3}+\frac{60}{\ee%
^4x^4}\rp  \; \frac{e^{-\ee x}}{x^3}\;, \label{3a.23}\\[2mm]
U_2^\PS(x) &=& \frac{1}{(4\p)^4}\; \frac{1}{6}\int_0^1\!d\a \int_0^1\!d\g \; %
\g\;\ee^4\; \lc  [3\!+\!(1\!-\!2\a)^2](1\!-\!2\g)^2 + 2 [1\!-\!(1\!-\!2\a%
)^2] (1\!-\!2\g) \right. \nn\\[2mm]
&+&\left.  [27\!+\!(1\!-\!2\a)^2]\rc
\lp 1+ \frac{10}{\ee x}+ \frac{45 }{\ee^2 x^2}+\frac{105}{\ee^3x^3}+\frac{105%
}{\ee^4 x^4} \rp \; \frac{e^{-\ee x}}{x^3}\;, \label{3a.24}\\[2mm]
U^\PV(x) &=&-\; \frac{1}{(4\p)^4}\; \frac{8}{3}\int_0^1\!d\a \int_0^1\!d\g %
\; \a(1-\a)\g^3(1-\g)\;\ee^5\; \nn\\
&\times& \lp 1+\frac{6}{\ee x}+\frac{15}{\ee^2x^2}+\frac{15}{\ee^3x^3}\rp \; 
\frac{e^{-\ee x}}{x^4} \label{3a.25} 
\eea

\ni and

\beq
\ee=\sqrt{\frac{1}{1-\g}+\frac{4}{\g[1-(1-2\a)^2]}}\;. \label{3a.26} 
\eeq

The structure of this result is similar to that of the OPEP, which is given
by

\bea
V^\p(r) &=& \frac{1}{3} \lp\frac{g\m}{2m}\rp^2 \frac{\m}{4\p}\bt^{(1)}\!\cdot%
\bt^{(2)} \lc \bs^{(1)}\!\cdot\!\bs^{(2)} \lp \frac{e^{-x}}{x} \rp 
+ S_{12}\lb \lp 1+\frac{3}{x}+\frac{3}{x^2} \rp  \frac{e^{-x}}{x}\rb\rc \;. 
\label{3.28} 
\eea

\begin{figure}[h]
\centerline{\epsfig{figure=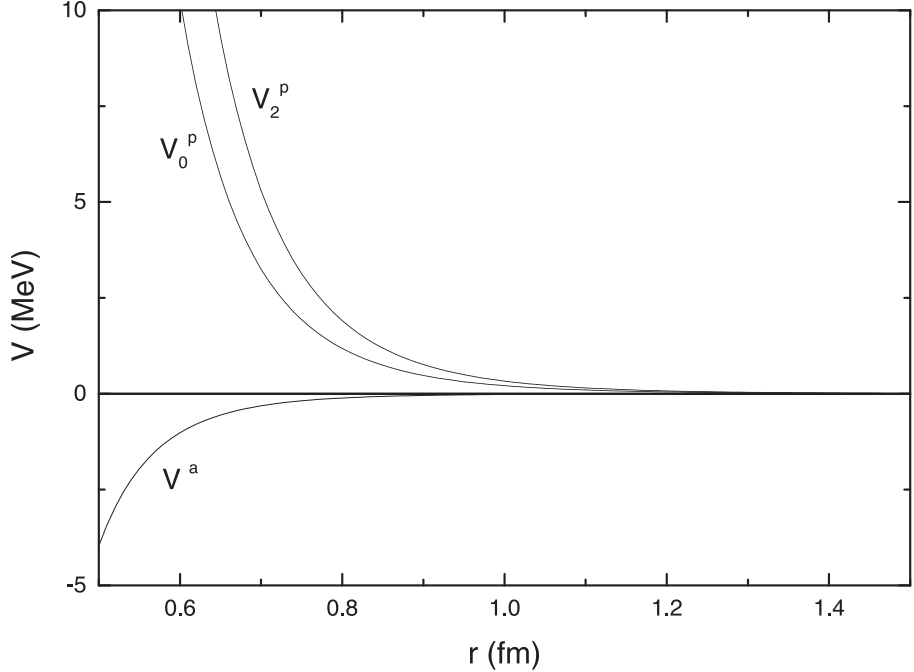,width=12.0cm}}
\caption{Profile function for the $V_0^p$, $V_2^p$ 
and $V^a$ component of the three-pion exchange potential.}
\end{figure}

The profile functions of the spin-spin and tensor components of the three-pion
exchange potential are displayed in fig.3, where it is possible to note that
all the curves show the typical divergent behaviour of unregularized potentials
at the origin. Therefore we assume that our results are realistic for
internucleon distances larger than 0.7 fm, the usual bag radius. Inspecting the
figure for the spin-spin chanel, one learns that the contribution of the axial
component is quite small and hence the three-pion exchange potential is
dominated by the pseudoscalar channel. In both $V_{SS}$ and $V_T$ its
contribution tends to add to the OPEP and is visible up to 1.5 fm, as shown in
fig.4. The influence of this component of the force over observables will be
discussed elsewhere.

\begin{figure}[h]
\centerline{\epsfig{figure=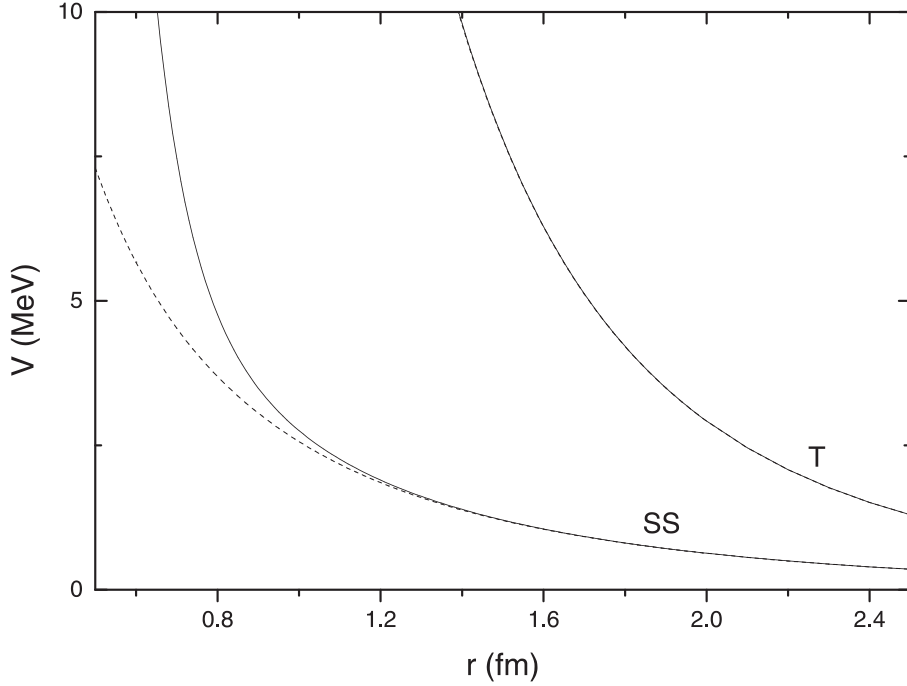,width=12.0cm}}
\caption{Profile function for the spin-spin (SS) and tensor (T) 
components of the $V^\pi$ (dashed line) and $V^\pi + V^{3\pi}$ (solid line).}
\end{figure}

In order to produce a feeling for the structure of the functions $U(x)$, in
the appendix we have evaluated approximately the integrals in eqs.(\ref
{3a.23}-\ref{3a.25}) and obtained the following asymptotic results ($x\rar %
\infty$)

\bea
U_0^\PS(x) &\rar& \frac{\p}{(4\p)^4}\; \frac{80}{\sqrt{3}} \lp 1+\frac{3}{x}+%
\frac{13}{3x^2}+\frac{10}{3x^3}+\frac{10}{9x^4}\rp \;\frac{e^{-3x}}{x^4}\;, 
\label{3.29}\\[2mm]
U_2^\PS(x) &\rar& \frac{\p}{(4\p)^4}\; \frac{80}{\sqrt{3}} \lp 1+\frac{4}{x}+%
\frac{20}{3x^2}+\frac{16}{3x^3}+\frac{16}{9x^4}\rp \;\frac{e^{-3x}}{x^4}\;, 
\label{3.30}\\[2mm]
U^\PV(x) &\rar& -\; \frac{\p}{(4\p)^4}\; \frac{16}{3\sqrt{3}} \lp 1+\frac{2}{%
x}+\frac{5}{3x^2}+\frac{5}{9x^3}\rp \;\frac{e^{-3x}}{x^5}\;, \label{3.31} 
\eea

\ni which are compared with the exact ones in fig.5.

\begin{figure}[h]
\centerline{\epsfig{figure=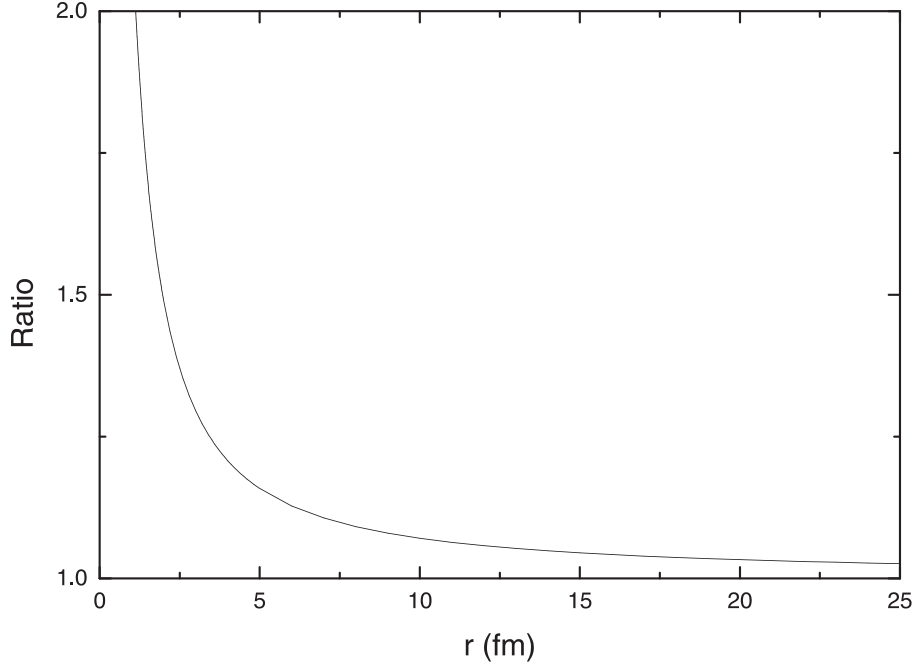,width=12.0cm}}
\caption{Ratios between the approximate expressions (\ref{3.29}-\ref{3.31})
by the corresponding functions (\ref{3a.23}-\ref{3a.25}) - all ratios are
identical.}
\end{figure}

A last point we would like to address here concerns the nature of the force
in the chiral limit. The potential given in eqs.(\ref{3.22}-\ref{3a.26})
incorporates two kinds of approximations. The first of them is associated
with the assumption that ${\bar{\mathcal{L}}}$, eq.(\ref{2.19}), represents
the leading contribution to the NN$\p\p\p$ vertex. The other one is related
to the non-relativistic limit taken in eq.(\ref{3.17}). On the other hand,
no approximations besides the neglect of contact interactions were performed
in the calculation of the three-pion propagator represented by the functions 
$I(\D)$. Therefore the corresponding configuration space expressions, given
by eqs.(\ref{3a.23}-\ref{3a.25}) also do not contain approximations and can
be used to evaluate the form of the interaction in the chiral limit. The
strength of $V^{3\pi }(r)$, as given by eq.(\ref{3.22}), is proportional to $%
\m^{7}$. Recalling that $x=\m r$, we obtain the following results when $\m%
\rar0$: $\m^{7}U_{0}^{\PS}(x)\rar140/[(4\p)^{4}\;r^{7}]$, $\m^{7}U_{2}^{\PS%
}(x)\rar245/[(4\p)^{4}\;r^{7}]$ and $\m^{7}U^{\PV}(x)\rar-5/[(4\p%
)^{4}\;r^{7}]$. Thus, the three-pion exchange NN potential survives in the
chiral limit, where it has the form

\bea
V^{3\p}(r) &\rar& \frac{1}{3} \lp\frac{g}{4m f_\p^2}\rp^2 \frac{1}{(4\p)^5}%
\; \bt^{(1)}\!\cdot\bt^{(2)} \lb 125\; \bs^{(1)}\!\cdot\!\bs^{(2)} + 245
\;S_{12} \rb\; \frac{1}{r^7} \;. \label{3.chi} 
\eea

%.............................................................................

\vs{10mm} 

\ni \textbf{acknowledgment}

J.C.P. would like to thank FAPESP for financial support.

%AAAAAAAAAAAAAAAAAAAAAAAAAAAAAAAAAAAAAAAAAAAAAAAAAAAAAAAAAAAAAAAAAAAAAAAAAAAAAA

\appendix

\section{Integrals}

In this appendix we evaluate the integral $I_{\m\n}$ given by eq.(\ref{3.13}%
), using the following results

\bea
&&X(K;\m,\x) = \int \frac{d^4 Q}{(2\p )^4}\; \frac{1}{[(Q\!-\!K/2)^2\!-\!\m%
^2]\;[(Q\!+\!K/2)^2\!-\!\m^2\x^2]} \nn\\[2mm]
&&= \frac{i}{(4\p)^2}\int_0^1\!d\a \lb \r_0\!-\!\ln\lp 1\!- \frac{K^2}{\m^2
\S^2} \rp\rb \;, \label{a.1}\\[2mm]
&&X_\m(K;\m,\x) = \int \frac{d^4 Q}{(2\p )^4}\; \frac{Q_\m }{%
[(Q\!-\!K/2)^2\!-\!\m^2]\;[(Q\!+\!K/2)^2\!-\!\m^2\x^2]} \nn\\[2mm]
&&= -\; \frac{i}{(4\p)^2} K_\m \int_0^1\!d\a \lp \frac{1-2\a}{2}\rp \lb %
\r_0\!-\!\ln\lp 1\!-\frac{K^2}{\m^2 \S^2} \rp\rb \;, \label{a.2}\\[2mm]
&&X_{\m\n}(K;\m,\x) = \int \frac{d^4 Q}{(2\p )^4}\; \frac{Q_\m Q_\n}{%
[(Q\!-\!K/2)^2\!-\!\m^2]\;[(Q\!+\!K/2)^2\!-\!\m^2\x^2]} \nn\\[2mm]
&&= \frac{i}{(4\p)^2} \lc K_\m K_\n \int_0^1\!d\a \lp \frac{1-2\a}{2}\rp^2 %
\lb \r_0\!-\!\ln\lp 1\!-\frac{K^2}{\m^2 \S^2} \rp\rb
\right. \nn\\[2mm]
&& +\left. \frac{\m^2}{2}\; g_{\m\n} \int_0^1 \!d\a \;\frac{ \S^2
[1\!-\!(1\!-\!2\a)^2]}{4}\lp 1\!-\frac{K^2}{\m^2\S^2} \rp 
\lb \r_1\!-\!\ln\lp 1\!- \frac{K^2}{\m^2\S^2} \rp\rb \rc \;, \label{a.3} 
\eea

\ni where

\beq
\S^2 = \frac{4[\a+(1-\a)\x^2]}{[1-(1-2\a)^2]} \;, \label{a.4} 
\eeq

\ni $\r_0$ and $\r_1$ are constants associated with the
dimensional regularization procedure. In order to perform the integrations,
it is convenient to use the following representation for the logarithm

\beq
\ln \lp 1-\frac{K^2}{\m^2\S^2}\rp 
= \int_0^1\! d\b\;\lp \frac{1}{\b}+ \frac{\m^2\S^2 / \b^2}{K^2 - \m^2\S^2 /
\b}\rp \;. \label{a.5} 
\eeq

Quite generally, constants appearing in these results correspond to contact
interactions, since they do not depend on $\D$. As we are interested in the
long range part of the potential, these constants will be neglected in the
sequence and we write

\bea
X(K;\m,\x) &=& -\;\frac{i}{(4\p)^2}\int_0^1\!d\a 
\int_0^1\!d\b \;\frac{\m^2\S^2/\b^2}{K^2\!-\!\m^2\S^2/\b} \;, \label{a.6}\\[%
2mm]
X_\m(K;\m,\x) &=& \frac{i}{(4\p)^2} \int_0^1\!d\a \int_0^1\!d\b \;\frac{\m%
^2\S^2/\b^2}{K^2\!-\!\m^2\S^2/\b} \lb \lp \frac{1\!-\!2\a}{2}\rp K_\m \rb %
\;, \label{a.7}\\[2mm]
X_{\m\n}(K;\m,\x) &=& -\; \frac{i}{(4\p)^2} \int_0^1\!d\a  \int_0^1\!d\b \;%
\frac{\m^2\S^2/\b^2}{K^2\!-\!\m^2\S^2/\b} \lc \lp \frac{1\!-\!2\a}{2}\rp^2
K_\m K_\n \right. \nn\\[2mm]
&-& \left. \lb \frac{ \S^2 [1\!-\!(1\!-\!2\a)^2] (1\!-\!\b)}{ 8\b}\rb \;\m%
^2\; g_{\m\n} \rc \;. \label{a.8} 
\eea

The integral in $Q^{\prime}$ can be performed using these results and one has

\bea
I'_{\m\n} &=& \int \frac{d^4 Q'}{(2\p)^4}\; \frac{[3Q_\m Q_\n\!-\!Q_\m\D%
_\n/2\!-\!\D_\m Q_\n/2\!+\!27\D_\m\D_\n /4+4Q'_\m Q'_\n]} {[(Q'\!-\!Q/2\!-\!%
\D/4)^2\!-\!\m^2][(Q'\!+\!Q/2\!+\!\D/4)^2\!-\!\m^2]} \nn\\[2mm]
&=& (3Q_\m Q_\n\!-\!Q_\m\D_\n/2\!-\!\D_\m Q_\n/2\!+\!27\D_\m\D_\n /4)/ \m%
^2\; X(Q\!+\!\D/2;\m,1) \nn\\[2mm]
&+& 4\;X_{\m\n}(Q\!+\!\D/2;\m,1) \nn\\[2mm]
&=& -\; \frac{i}{(4\p)^2}\;\m^2\; \int_0^1\!d\a \int_0^1\!d\b\;\frac{\mb^2}{%
\b}\; \frac{1}{[(Q\!+\!\D/2)^2\!-\!\m^2\mb^2]} \lc [3\!+\!(1\!-\!2\a)^2]%
\;Q_\m Q_\n \right. \nn\\[2mm]
&-& \left. [1\!-\!(1\!-\!2\a)^2](Q_\m \D_\n \!+\! \D_\m Q_\n)/2
+[27\!+\!(1\!-\!2\a)^2]\; \D_\m \D_\n/4 \right. \nn\\[2mm]
&-& \left. \mb^2 [1\!-\!(1\!-\!2\a)^2](1\!-\!\b)\;\m^2\; g_{\m\n}/2 \rc \;, 
\label{a.9} 
\eea

\ni where

\beq
\mb^2 = \frac{4}{\b [1-(1-2\a)^2]}\; . \label{a.10} 
\eeq

The function $I_{\m\n}$ is then given by

\bea
&& I_{\m\n}(\D) =\frac{1}{3!}\; \int \frac{d^4Q}{(2\p)^4}\; \frac{1}{[(Q-\D%
/2)^2\!-\!\m^2]}\;I'_{\m\n} \nn\\[2mm]
&&= -\;\frac{i}{(4\p)^2} \frac{1}{3!}\; \m^2 \int_0^1\!d\a \int_0^1\!d\b\; 
\frac{\mb^2}{\b} \lc [3\!+\!(1\!-\!2\a)^2]\; X_{\m\n}(\D,\m,\mb) \right. \nn%
\\[2mm]
&&- \left. [1\!-\!(1\!-\!2\a)^2]\;[\D_\m X_\n(\D,\m,\mb)\!+\!\D_\n X_\m(\D,\m%
,\mb)]/2 +[27\!+\!(1\!-\!2\a)^2] X(\D,\m,\mb)\;\D_\m\D_\n/4 \right. \nn\\[2mm%
]
&&-\left. \mb^2 [1\!-\!(1\!-\!2\a)^2] (1\!-\!\b)\; X(\D,\m,\mb)\;\m^2 \;g_{\m%
\n}/2 \rc
\nn\\[2mm]
&& = \D_\m\D_\n\; \m^4\;I^\PS(\D) + g_{\m\n}\;\m^6\;I^\PV(\D) \;, \label%
{a.11} 
\eea

\ni where

\bea
I^\PS(\D) &=& -\;\frac{1}{(4\p)^4}\; \frac{1}{24}\int_0^1\!d\a%
\int_0^1\!d\b\int_0^1\!d\g\int_0^1\!d\e\; \frac{\mb^2 \th^2}{\b\e} \lc 
[3\!+\!(1\!-\!2\a)^2](1\!-\!2\g)^2 \right. \nn\\[2mm]
&+&\left. 2 [1\!-\!(1\!-\!2\a)^2] (1\!-\!2\g) + [27\!+\!(1\!-\!2\a)^2]\rc \; 
\frac{1}{[\D^2\!-\!\m^2\th^2]}\;, \label{a.12}\\[2mm]
I^\PV(\D) &=& \frac{1}{(4\p)^4}\; \frac{1}{48}\int_0^1\!d\a%
\int_0^1\!d\b\int_0^1\!d\g\int_0^1\!d\e\; \frac{\mb^2 \th^2}{\b\e} \lc 4\;\mb%
^2 \;[1\!-\!(1\!-\!2\a)^2](1\!-\!\b) \right. \nn\\[2mm]
&+& \left. \th^2 \; [3\!+\!(1\!-\!2\a)^2][1\!-\!(1\!-\!2\g)^2] (1\!-\!\e) %
\rc  \; \frac{1}{[\D^2\!-\!\m^2\th^2]}\;, \label{a.13} 
\eea

\ni with

\beq
\th^2= \frac{4[\g+(1-\g)\mb^2]}{\e[1-(1-2\g)^2]}\;. \label{a.14} 
\eeq

Results presented in this appendix are covariant. On the other hand, the
calculation of the potential is performed in the centre of mass of the NN
system and we use $\D=(0;\bD)\rar\D^2=-\bD^2$.

The potential in configuration space is determined by the functions $U(x)$,
given by

\beq
U(x) = \frac{4\p}{\m} \int \frac{d^3 \bD}{(2\p)^3}\; e^{-i\bD \cdot \br}\; I(%
\D) \;. \label{a.15} 
\eeq

\ni where $x\equiv \m r$. Using the result

\beq
\int \frac{d^3 \bD}{(2\p)^3}\; \frac{e^{-i\bD \cdot \br}}{\bD^2+\m^2\th^2} =%
\frac{\m}{4\p} \; \frac{e^{-\th x}}{x}\;, \label{a.16} 
\eeq

\ni we have

\bea
U^\PS(x) &=& \frac{1}{(4\p)^4}\; \frac{1}{24}\int_0^1\!d\a%
\int_0^1\!d\b\int_0^1\!d\g\int_0^1\!d\e\; \frac{\mb^2 \th^2}{\b\e} \lc 
[3\!+\!(1\!-\!2\a)^2](1\!-\!2\g)^2 \right. \nn\\[2mm]
&+&\left. 2 [1\!-\!(1\!-\!2\a)^2] (1\!-\!2\g) + [27\!+\!(1\!-\!2\a)^2]\rc \; 
\frac{e^{-\th x}}{x}\;, \label{a.17}\\[2mm]
U^\PV(x) &=&-\; \frac{1}{(4\p)^4}\; \frac{1}{48}\int_0^1\!d\a%
\int_0^1\!d\b\int_0^1\!d\g\int_0^1\!d\e\; \frac{\mb^2 \th^2}{\b\e} \lc 4\;\mb%
^2 \;[1\!-\!(1\!-\!2\a)^2](1\!-\!\b) \right. \nn\\[2mm]
&+& \left. \th^2 \; [3\!+\!(1\!-\!2\a)^2][1\!-\!(1\!-\!2\g)^2] (1\!-\!\e) %
\rc  \; \frac{e^{-\th x}}{x}\;. \label{a.18} 
\eea

The integrations in $\e$ and $\b$ can be performed analytically and we have

\bea
U^\PS(x) &=& \frac{1}{(4\p)^4}\; \frac{1}{6}\int_0^1\!d\a \int_0^1\!d\g \; \g%
\;\ee^2\; \lc  [3\!+\!(1\!-\!2\a)^2](1\!-\!2\g)^2 + 2 [1\!-\!(1\!-\!2\a)^2]
(1\!-\!2\g) \right. \nn\\[2mm]
&+&\left.  [27\!+\!(1\!-\!2\a)^2]\rc
\lp 1 +\frac{3}{\ee x}+\frac{3}{\ee^2x^2}\rp \; \frac{e^{-\ee x}}{x^3}\;, 
\label{a.19}\\[2mm]
U^\PV(x) &=&-\; \frac{1}{(4\p)^4}\; \frac{8}{3}\int_0^1\!d\a \int_0^1\!d\g %
\; \a(1-\a)\g^3(1-\g)\;\ee^5\; \nn\\
&\times& \lp 1+\frac{6}{\ee x}+\frac{15}{\ee^2x^2}+\frac{15}{\ee^3x^3}\rp \; 
\frac{e^{-\ee x}}{x^4}\;, \label{a.20} 
\eea

\ni where

\beq
\ee=\sqrt{\frac{1}{1-\g}+\frac{4}{\g[1-(1-2\a)^2]}}\;. \label{a.21} 
\eeq

The integrals over $\a$ and $\g$ can be evaluated approximately for large
values of $x$. In this case, the exponential has a sharp minimum for $\a%
=1/2, \g=2/3$ and varies very rapidly around it. Thus all the elements in
the integrand but the exponential may be taken as constants and we have

\bea
U^\PS(x) &=& \frac{1}{(4\p)^4}\; \frac{80}{3} \lp 1 +\frac{1}{x}+\frac{1}{%
3x^2}\rp \; \frac{1}{x^3} \int_0^1\!d\a \int_0^1\!d\g \;e^{-\ee x} \;, \label%
{a.22}\\[2mm]
U^\PV(x) &=&-\; \frac{1}{(4\p)^4}\; 16 \lp 1+\frac{2}{x}+\frac{5}{3x^2}+%
\frac{5}{9x^3}\rp \; \frac{1}{x^4} \int_0^1\!d\a \int_0^1\!d\g \; e^{-\ee %
x}\;. \label{a.23} 
\eea

In order to perform the last integral we first use a new variable $u$,
related to $\a$ by

\beq
\a = \frac{1}{2} - \lb \frac{1}{4}- \frac{1}{4 + 2\g u^2 \sqrt{\frac{4-3\g}{%
\g(1-\g)}} +\g u^4 } \rb^{1/2} \label{a.24} 
\eeq

\ni and then another variable $v$, related to $\g$ by

\beq
\g= \frac{3+\lp 3+v^2 \rp^2\mp \sqrt{48v^2+44v^2+12v^6+v^8}}{2 \lp 3+v^2 \rp%
^2}\;, \label{a.25} 
\eeq

\ni where the ($-$) and ($+$) signs refer to the intervals $0\leq\g\leq 2/3$
and $2/3\leq\g\leq1$ respectively. We then obtain

\beq
\int_0^1\!d\a \int_0^1\!d\g \;e^{-\ee x}=\frac{\p}{3\sqrt{3}}\; \frac{e^{-3x}%
}{x}\;. \label{a.26} 
\eeq

%RRRRRRRRRRRRRRRRRRRRRRRRRRRRRRRRRRRRRRRRRRRRRRRRRRRRRRRRRRRRRRRRRRRRRRRRRRRRRR

\end{document}